\pdfoutput=1
\documentclass[useAMS,usenatbib,usegraphicx]{mn2e}
\title[Straight segments  in the galactic discs]
{Straight segments in the galactic discs}\author[A.M. Mel'nik and
P. Rautiainen] {A.M. Mel'nik$^{1}$\thanks{E-mail:
anna@sai.msu.ru} and P. Rautiainen$^{2}$ \\
$^{1}$ Sternberg Astronomical Institute, Lomonosov Moscow State
University,\\ Universitetskij pr. 13, Moscow 119991, Russia\\
$^{2}$ Astronomy Division, Department of Physics, University of
Oulu, P.O. Box 3000,\\ FI-90014 Oulun yliopisto, Finland}

\begin{document}

\date{Accepted 2013 June 16. Received 2013
June 3; in original form 2013 April 5}
\pubyear{2012}

\maketitle

\label{firstpage}

\begin{abstract}
We study the properties of the straight segments forming in
N-body simulations of the galactic discs. The properties of these
features are consistent with the observational ones summarize by
Chernin at al. (2001). Unlike some previous suggestions to
explain the straight segments as gas dynamical instabilities,
they form in our models in the stellar system. We suggest that
the straight segments are forming as a response of the rotating
disc to a gravity of the regions of enhanced density
(overdensities) corotating with the disc. The kinematics of stars
near the prominent overdensities is consistent with this
hypothesis.
\end{abstract}

\section{Introduction}

Straight segments in  spiral structure of galactic discs are
observed both in real galaxies and in numerical models. The
straight segments  were first noticed by \citet{vorontsov1964,
vorontsov1978}, who called them rows. These quite long nearly
straight features often outline the grand design spiral
structure, as it is, for example, in M101 and in M51, forming
ragged but nearly regular spiral arms which are often called
polygonal arms. Chernin at al. (Chernin, Zasov, Arkhipova, \&
Kravtsova, 2000;  Chernin, Kravtsova, Zasov, \& Arkhipova 2001)
compiled the catalog of galaxies with rows that includes about
200 objects. They also study the properties of straight segments,
which can be briefly formulated as follows.

\begin{enumerate}

\item The length of the straight segment $L$ increases with the
galactocentric distance $R$, so that $L=(1\pm 0.13) R$.

\item  The straight segments can be divided into two types: those
that   fit well the grand design spiral arms and isolated ones.

\item The angle between two neighboring segments is, on average,
$\alpha=120$, the standard deviation is $\sim 10^\circ$.

\item The straight segments are observed mostly in late type
galaxies Sbc-Scd.

\item The straight segments  are more frequently observed in
interacting galaxies.

\item The average number of straight segments in the polygonal
spiral arms is $n=3$.

\item Galaxies with straight segments are quite rare objects,
they account for $\sim7$ per cent of all spiral galaxies with
well-defined spiral arms. Note however that this estimate is
based on studying photographic plates and printed images. Our
inspection of a small sample of digital galaxy images suggests a
considerably larger frequency of straight segments in spiral
galaxies.

\end{enumerate}

Straight segments in numerical models are observed quite
frequently in both gaseous and stellar discs, but their
appearance is often mentioned only briefly  because the emphasis
of the authors have   been on larger scale structure.

The straight segments in gaseous discs are observed in models by
\citet{combes1994}, who studies the gas inflow  in barred
potentials. She explains the  square-like shape of the spiral
arms by the resonances and gas viscosity: the periodic orbits
must change their orientation with respect to the bar in the
Inner and Outer Lindblad resonances: ILRs, OLR \citep{buta1996}.

Khoperskov, Khoperskov, Eremin, \& Butenko (2011) and
\citet{filistov2012} study the formation of straight segments in
the gaseous discs under the given analytical  potential. Their
models include the external spiral-like potential perturbation,
which rotates with a small angular velocity. As they note, the
position of the corotation radius (CR) on the very periphery of
the disc is a necessary condition  for the appearance of straight
segments in their models. Simulations by \citet{khoperskov2011}
reproduce well the main properties of the straight segments: the
dependence $L\sim R$ and the average angle between the
neighboring segments $\alpha=120^\circ$. They explain the
formation of the straight segments by unstable location of the
shock fronts in the spiral potential well.

\citet{chernin1999} explains the formation of straight segments
as the universal stability of a flat shock front against any weak
perturbations that disturb its front surfaces.
\citet{filistov2012}  also supports this idea. But it is not
clear how this mechanism works in the rotating stellar systems
\citep[][for more comments]{khoperskov2011}.

Rautiainen, Salo, \& Laurikainen (2005, 2008), using the
potentials extracted from the near-IR images  of Ohio State
University Bright Spiral Galaxy Survey \citep[][OSUBSGS,
hereafter]{eskridge2002}, model the behavior of the gas subsystem
of some disc galaxies. Their models reproduce the straight
segments observed in some galaxies, especially well in the case
of NGC 4303.

\begin{figure*}
\centering \resizebox{\hsize}{!}{\includegraphics{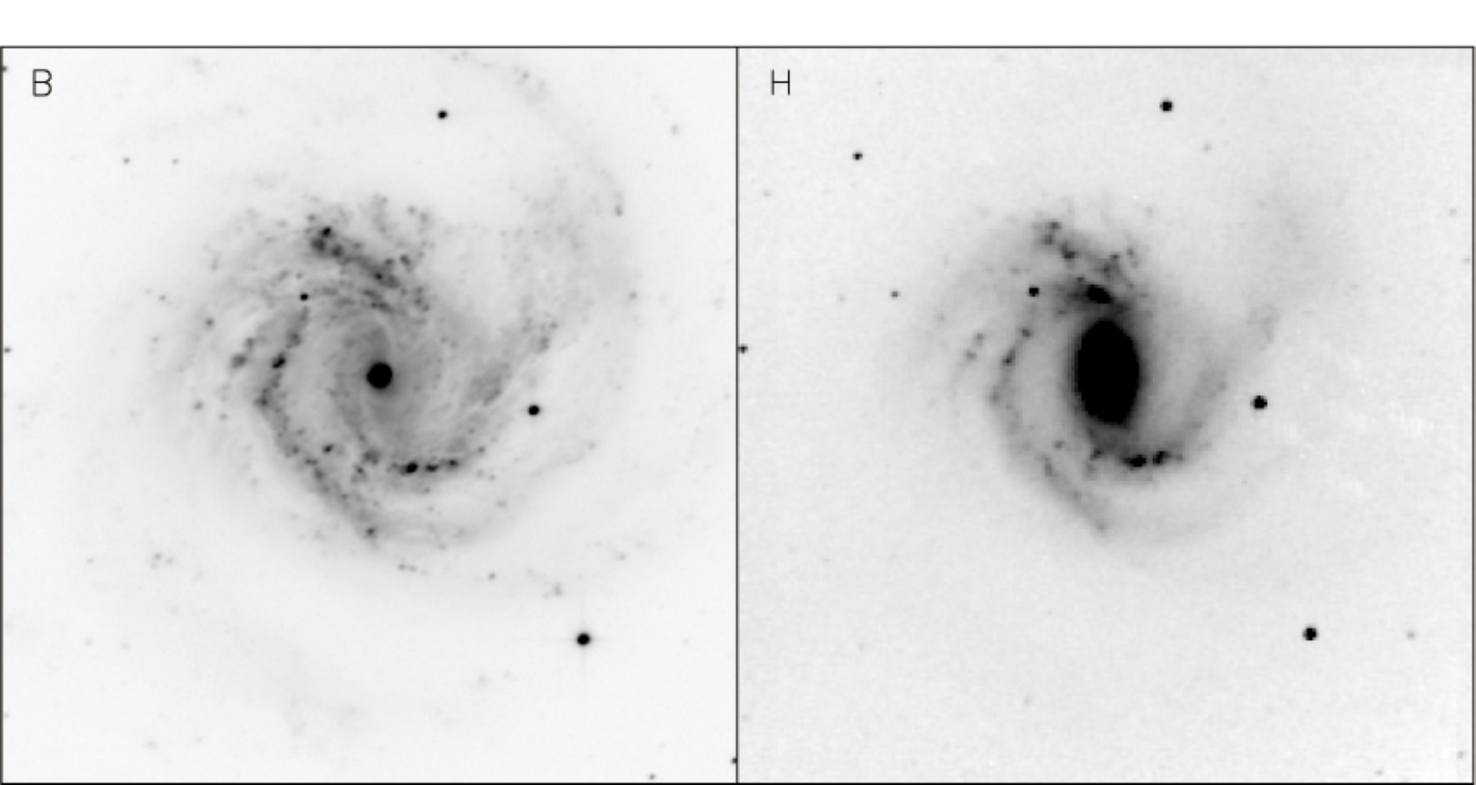}}
\caption{The B- and H-band images of NGC 4303. The images are
taken  from the   Ohio State University Sample of Bright Galaxies
\citep{eskridge2002}, and have been scaled to enhance the
visibility  of the straight segments.} \label{OSUSBG_4303}
\end{figure*}

The successful modelling of NGC 4303 is at least partly based on
the fact that the straight segments are observed also in the
near-infrared H-band image, which should be dominated by the old
stellar population (see Fig.~\ref{OSUSBG_4303}), i.e. they were
present in the derived gravitational potential. Furthermore, NGC
4303 is not an exceptional case. We found about 40 galaxies with
straight segments in the OSUBSGS images: their overall frequency
in the sample was $\sim$ 25 per cent , considerably higher than
$\sim$ 7 per cent found in earlier studies.  This difference was
most likely due to advantage of the digital images  over the
photographic plates and atlases used by \citet{chernin2001} --
the possibility to adjust contrast and other image properties
revealed many straight segments that were  missed in ``static
images''. Our inspection also revealed that in more than half of
the cases, the straight segments observed in B-band had their
counterparts also in H-band. In addition to our findings, the
straight segments are quite conspicuous in near-infrared J- and
K-band images of galaxies NGC 3938 and NGC 4254 obtained by
\citet{castro-rodriguez2003}.

There are two different approaches to  explain the formation of
straight segments in the stellar subsystem: one is based on the
global modes \citep{toomre1981} and the other rests on the
chaotically distributed rotating features \citep{toomre1991}.

The most popular  explanation of the polygonal spiral arms is
proposed by \citet{toomre1981}. He explains the square-like shape
of the spiral arms by the presence of the leading and trailing
spiral waves of very similar wavelengths and amplitudes in the
Fourier spectrum of the mode, where the leading wave appears due
to reflection of the in-going trailing wave from the centre
\citep[][for more details]{athanassoula1984, binney2008}.

\citet{salo2000b} explain  the inner polygonal structure of M 51
by  the reflection of the trailing wave packets as leading waves
from the centre. Their simulations reproduce the polygonal spiral
arms in the inner 30 arcsec region of M 51 observed in near-IR
\citep{zaritsky1993}.

However, it is not clear how to produce the superposition of the
leading and trailing waves on the galactic periphery.   In this
context, it is worth noting ideas  by \citet{sellwood2012}, who
supposes that the region of the ILR can acquire ability to
reflect the in-going trailing waves into the out-going leading
ones.

In the other approach the observed spiral structure is considered
as a set of arm features  forming due to  random density
fluctuations in galactic discs \citep{toomre1990}.
\citet{julian1966} consider the response of the stellar disc to a
chance overdensity corotating with the disc. The  density
response  can exceed the initial perturbation more than several
tens of times \citep{goldreich1965, julian1966, toomre1981}. This
mechanism called swing amplification  is based on the concerted
action of noise, epicyclic motion, and self-gravity
\citep{toomre1981}. \citet{sellwood1984} study the work of the
swing amplification mechanism and show that the maximal
amplification is possible on the galactic periphery for the
multi-armed spiral patterns.

Recently, many researchers note that the multi-armed spiral
structure in their N-body simulations doesn't rotate as a whole,
but consists of pieces corotating with the disc  at different
radii (Wada, Baba, \& Saitoh 2011; Grand, Kawata, \& Cropper
2012; Baba, Saitoh, \& Wada 2013; D'Onghia, Vogelsberger, \&
Hernquist 2013; Roca-F\`abrega et al. 2013).

\citet{d_onghia2013} study stellar discs with the TreePM code
GADGET-3 using  small softening parameter ($\epsilon=5$ pc). They
get very impressive pictures of polygonal spiral arms (or linear
segments joined at kinks), which form global multi-armed spiral
stricture. In their experiments the system of disturbers
($M\approx10^6$ M$_\odot$) corotating with the disc causes the
formation of the multi-armed polygonal spiral arms.

\citet{grand2012} study the motions of stars near the spiral arms
in N-body simulations. Their stellar discs form multi-armed
structures,  which often exhibit straight segments. They show
that particles can join spiral arms from both sides at all radii
and migrate radially along the spiral arms.

In the present paper we  study  properties of the straight
segments forming in N-body galactic discs. We show that the
features of the model straight segments are in a good agreement
with the observational ones summarized by \citet{chernin2000,
chernin2001}. We suppose that the straight segments are forming
as a response of the rotating disc to a gravity of the regions of
enhanced density (overdensities) corotating with the disc. The
properties of these respondent perturbations can explain the
observational features of the straight segments. The kinematics
of stars in the model discs also agrees with this suggestion.

In section 2 we study   kinematical and morphological properties
of the respondent perturbation of the disc to the overdensity
co-rotating with it, that was  first studied by
\citet{julian1966}. We show that the respondent perturbation must
be nearly straight and its length $L$ is nearly proportional to
$R$. The models and their evolution are considered in section 3.
Here we also demonstrate that the model spiral pattern doesn't
rotate with the same angular velocity, but its different parts
nearly corotate with the disc. In section 4 we compare the
characteristics of the model straight segments with the
theoretical predictions studied in Section 2. Section 5 is
devoted to the kinematics of the straight segments. We compare
the stellar motions near the overdensities with  the theoretical
predictions. Section 6 includes  the main conclusions.

\section{The properties of the disc response to the
overdensity}

\subsection{The shape of the respondent perturbation}

\citet{toomre1964} has shown that the stability of the disc is
supported by shared action of  the Coriolis forces  and the
equivalent of pressure, resulting from random motions: the random
motions effectively suppress perturbations on the short side of
wavelengths, while the Coriolis forces suppress instabilities on
the long end. The value of $\lambda_c$ is the shortest wavelength
of axisymmetric perturbations that can be stabilized by epicyclic
motions only:

\begin{equation}
  \lambda_c=\frac{4\pi^2G\Sigma}{\kappa^2},
  \label{lambda}
\end{equation}

\noindent where  $\Sigma$ is the surface density of the disc and
$\kappa$ -- the epicyclic frequency.

\citet{julian1966} consider the response of a thin differentially
rotating stellar disc to the presence of a single, particle-like
concentration of the interstellar matter (overdensity) corotating
with the disc. They have found that overdensity creates quite
extended spiral-like response in the disc: the size of the
density ridge in the radial direction amounts $\sim \lambda_c/2$.

\citet{toomre1981} studies the self-gravitating stellar discs
with flat rotation curves and shows that the value of
amplification of the initial overdensity  is very sensitive to
the value of the stability parameter $Q$ \citep{toomre1964}. The
other parameter that determines amplification is
$X=\lambda_y/\lambda_c$, where $\lambda_y$ is the length between
the neighboring spirals in the azimuthal direction. Maximum
amplification corresponds to $X\approx1.5$.

\begin{figure*}
\centering \resizebox{12 cm}{!}{\includegraphics{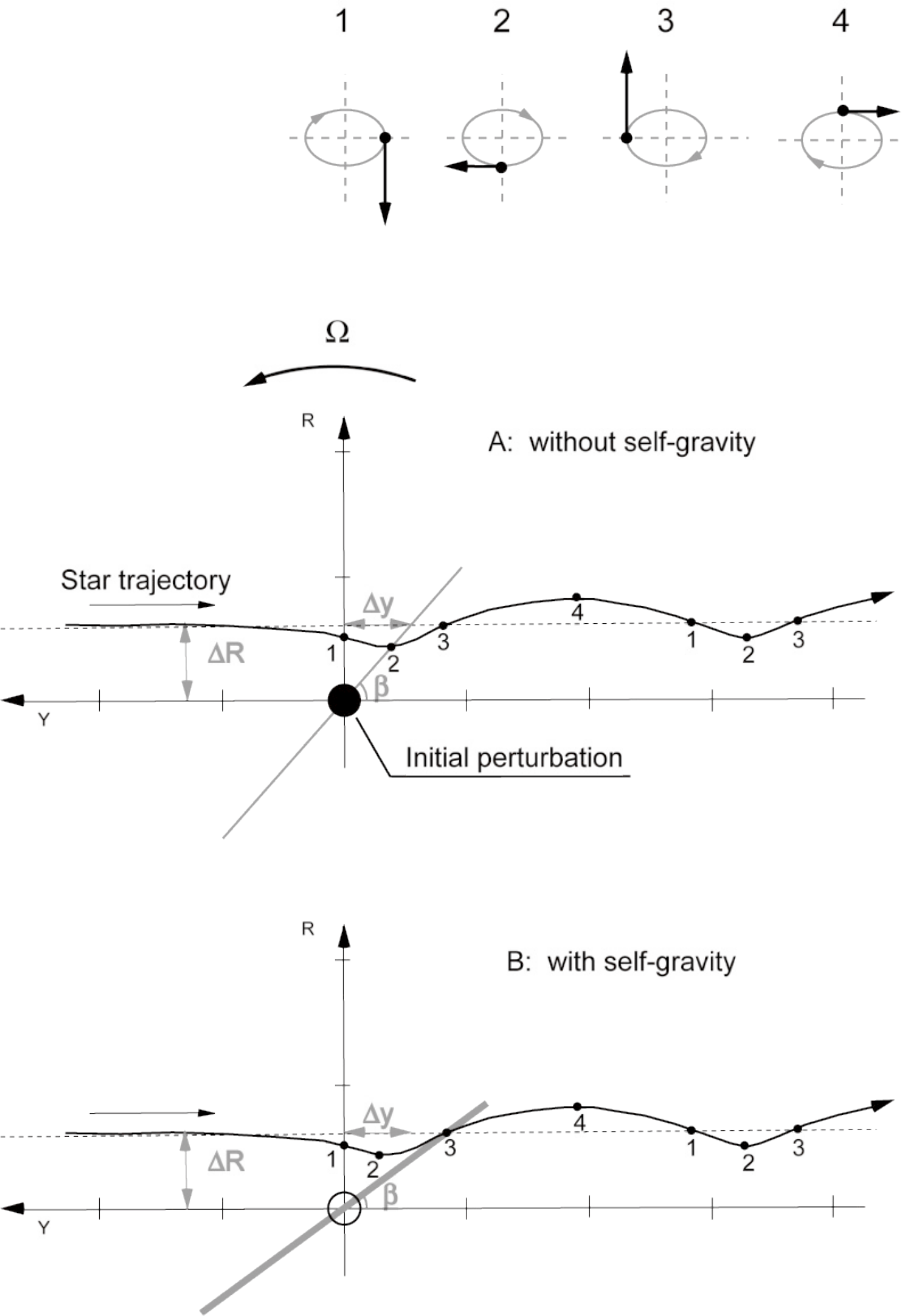}}
\caption{The trajectory of the star (black curve) perturbed by
the initial overdensity \citep{julian1966}. The motion is
considered in the reference frame, corotating with the initial
disturber, which lies at the origin and rotates with the circular
velocity. The star in question is lying at the distance larger
than that of the disturber and initially moves with purely
circular velocity. In the chosen reference frame, it moves in the
sense opposite that of galactic rotation.  The numbers 1--4
denote positions of the star at moments separated by 1/4 of the
epicyclic period. The upper row shows the position of the star in
the epicyclic orbit at moments 1-4. A: Position of the straight
segment without taking its self-gravity into account. Here the
greatest stellar density corresponds to the point "2". B:
Position of the straight segment with self-gravity. In this case
the highest density correspond to the point "3", where stars are
moving nearly along the straight segment.} \label{trajectory-1}
\end{figure*}

Fig.~\ref{trajectory-1}a shows the trajectory of a star with
respect of the initial overdensity  \citep{julian1966}. The star
in question is located at the larger distance than the disturber,
and first  has purely circular velocity. In the reference frame
corotating with the disturber, the star moves in the direction
opposite that of galactic rotation, i. e. clockwise. In the
impulse approximation, the star's angular momentum is unchanged
and its motion can be thought as a superposition of the purely
circular motion and the motion along the epicycle
\citep{binney2008}. Let us suppose that the star gains some
impulse and starts its epicyclic motion when the distance between
the star and the disturber is minimal, i.e. when the star and the
disturber are lying at the same radius-vector. The moment of
start of the epicyclic motion is denoted by number "1" and
corresponds to the  maximal additional velocity directed toward
the galactic centre. \citet{julian1966} suggest that the
resulting stellar density must be the greatest wherever the
individual stars linger longest. That moment denoted by number
"2" occurs in nearly one-quarter of the epicyclic period, when
the star has the maximal additional velocity directed in the
sense of galactic rotation. Note that in the chosen reference
frame, the star in question moves in the direction opposite that
of galactic rotation, so the moment with the largest velocity in
the sense of galactic rotation determines the place where the
star lingers most.  Moment "3" corresponds to the maximal
additional velocity directed away from the galactic centre. The
additional velocity in moment "4" is directed in the sense
opposite that of galactic rotation. In the absence of occasional
perturbations stellar trajectories must repeat their oscillations
every epicyclic period.

Let us calculate an angle $\beta$, which determines the position
of the respondent perturbation with respect to the azimuthal
direction and corresponds to the pitch angle of the spiral arms.
\citet{julian1966} suppose that without taking into account the
self-gravity  the angle $\beta$ must be $\sim 45^\circ$ for flat
rotation curve. Generally, the fact that the angle $\beta$ is
independent from the coordinates $\Delta R$ and $\Delta y$
suggests that the stellar response has the shape of a straight
line. We will show that the distance $\Delta y$ to the point,
where the  star linger most, is nearly proportional to $\Delta
R$. The angle $\beta$ can be determined by the ratio:

\begin{equation}
  \tan\beta=\frac{\Delta R}{\Delta y}.
  \label{beta}
\end{equation}

In the first approximation, we neglect the additional velocities
due to the epicyclic motions.  Then the  distance $\Delta y$,
which is passed by the star with respect to the initial disturber
during one-quarter of the epicyclic period $\pi/(2\kappa)$, is
determined by the relation:

\begin{equation}
  \Delta y=|\Omega(R_1)- \Omega(R_0)|\frac{\pi R_0}{2\kappa_1},
  \label{dy-1}
\end{equation}

\noindent where $\Omega(R)$ is the angular velocity of  rotation
curve. Subscripts  "1" and "0" are related to the star considered
and to the initial disturber, respectively. For  flat rotation
curve ($\Omega(R)=V_0/R$, $\kappa=\sqrt2\Omega$) we can express
the distance $\Delta y$ in the following form:

\begin{equation}
 \Delta y=\frac{\pi}{2\sqrt2}\Delta R,
  \label{dy-2}
\end{equation}

\noindent And the value of the angle $\beta$ is determined by the
expression:

\begin{equation}
  \beta=\arctan\frac{2\sqrt2}{\pi}=42^\circ,
  \label{beta}
\end{equation}

\noindent which is very close to the value suggested by
\citet{julian1966}. Thus, in the impulse approximation the value
of $\beta$ is independent from $\Delta R$, and the respondent
perturbation must have the shape of the straight segment.
However, the impulse approximation isn't accurate, especially in
the very vicinity of the initial disturber, because any star
changes its angular momentum during the approach phase of the
encounter  and passes the disturber with a slower relative
velocity than it would be without interaction \citep{julian1966}.

Note that near  the disturber, stars oscillate conspicuously in
the radial direction, moving first toward the disturber and then
away from it. And  the star can continue its radial oscillations
as it  moves  in the azimuthal direction.

\subsection{Influence of self-gravity}

In the  cold discs  (but $Q>1$) the self-gravity plays important
role,  so after some moment, the stellar trajectories are rather
determined by the gravity of the straight segment itself than by
the initial disturber.

Let us again consider the motion of the star initially moving on
the circular orbit  (Fig.~\ref{trajectory-1}b). The self-gravity
effects are maximal at the time interval, when the star is
leaving the straight segment and is moving nearly parallel to it.
In Fig.~\ref{trajectory-1}b it is a path between points "2" and
"3". Due to the gravity of the straight segment, the position of
the density maximum is shifting in the direction of the point
"3", because here the star has maximal value of the radial
velocity, which allows it to move along the the straight segment
during the longest period of time.

Using the approach described above, we can calculate the pitch
angle of the self-gravitating straight segment,   which must be
nearly two times less than the angle calculated without
self-gravity, because the time interval to reach the point "3" is
nearly two times larger ($\pi/\kappa$) than that needed to reach
the point "2" from the start of the epicyclic motion at the point
"1". In the first approximation, the angle $\beta$ of the
self-gravitating straight segment   equals:

\begin{equation}
  \beta=\arctan\frac{\sqrt2}{\pi}=24^\circ.
  \label{beta}
\end{equation}

This result agrees with the estimate by \citet{toomre1981}, who
thinks that self-gravity must decrease the value of the pitch
angle by two times at least. Moreover, he supposes that due to
the distortion of epicycle motions the pitch angle can drop to
15$^\circ$.

Note that the direction of the radial velocity $V_R$ inside the
self-gravitating straight segments  coincides with that in the
density-wave spiral arms \citep{lin1969}: at the larger R from
the initial disturber (outside the CR) stars located inside the
straight segment (inside the spiral arm) have the radial velocity
$V_R$ directed away from the galactic centre, while stars located
at the smaller R than the initial disturber (inside the CR)  have
the velocity $V_R$ directed toward the galactic centre.
Generally, on the edges of the straight segment stars must move
in the opposite directions away from each other (away from the
initial disturber).

\subsection{The length of the straight segments}

The most interesting parameter is the linear  size of the
respondent density perturbation. \citet{julian1966} shows that
the size  of the density ridge  in radial direction is $\Delta
R\approx\lambda_c/2$.

Let us compare the value of $\lambda_c$ with the radius $R$, at
which it is calculated \citep[see also Fig. 5a in][]{toomre1977}.
We can approximate the distribution of the disc density using the
relation:

\begin{equation}
  \Sigma\approx\frac{f_d V_0^2}{2\pi G R},
  \label{sigma}
\end{equation}

\noindent where  $V_0$ is the velocity of the rotation curve and
$f_d$ is the contribution of the disc to the total rotation
curve. This formula is absolutely true for Mestel's discs, but
for exponential discs it is true within several per cent
\citep{binney2008}. Substituting $\Sigma$ in Eq.~1 and using the
relations $\Omega(R)=V_0/R$ and $\kappa=\sqrt2\Omega$ for  flat
rotation curves we obtain:

\begin{equation}
\lambda_c\approx \pi f_d R.
\end{equation}

\noindent And the maximal size of the straight segment in the
radial direction $\Delta R$ is defined by the expression:

\begin{equation}
\Delta R\approx\pi f_d R/2.
\end{equation}

\noindent Then the full size  of the straight segment $L$, under
the angle  $\beta\approx42^\circ$, must be following:

\begin{equation}
L \approx 2.4 f_d R.
\end{equation}

\noindent Generally, the self-gravity effects cannot increase the
length of the  straight segments.

In the distance range considered in our models, the value of
$f_d$  varies in the limits $f_d=0.2-0.5$. So the maximal
possible length of the straight segment  must lie in the range
$L=(0.5\textrm{--}1.2) R$. On the whole, this result is
consistent with the observations \citep{chernin2001}.

\subsection{Amplitude of the velocity perturbation}

\citet{binney2008}, using the impulse approximation, estimate the
value of the radial velocity $V_R$ acquired by a star after the
encounter with a molecular cloud:

\begin{equation}
V_R=-\frac{Gm}{A_0 b^2},
\end{equation}

\noindent where $m$ is the mass of molecular cloud, which
corotates with the disc at the radius $R_c$  and $A_0$ is Oort
constant at this radius. The star considered is initially moving
on the circular orbit of radius $R$, so $b=R-R_c$ is the impact
parameter.

We can see that the value of acquired velocity $V_R$ depends on
the current value of Oort constant $A$, which varies with radius.
For  flat rotation curve, $A$ is inversely proportional to $R$,
$A=\Omega/2\sim 1/R$. So the overdensity of the same mass can
create the larger velocity perturbation on the galactic periphery
than in the intermediate regions. And the physics of this
dependence is clear: the smaller $A$, the weaker differential
rotation, the smaller relative velocity of passage, the more time
of  gravitational interaction. So on the galactic periphery stars
can acquire larger radial velocities after encounters with the
same overdensities.

\section{Models}

The N-body simulations used in this article were done by P.
Rautiainen during year 2012 by applying the code written by H.
Salo. In these  2D models we use a logarithmic polar grid with
216 azimuthal and 288 radial cells to calculate the gravitational
forces and    motions with leap-frog integration. The stellar
disc consists of 4 10$^6$ self-gravitating particles, but the
bulge and halo are analytical. The gas component is omitted in
this article, in models we do not show here, the gas component
was modelled as inelastically colliding massless test particles.
For more details on the code,  see \citet{salo1991} and
\citet{salo2000a}.

We have made a large set of models to study the formation and
evolution of straight segments in the galactic discs. In these
models we varied several parameters such as the mass fractions of
different components, the value of the initial Toomre-parameter
of the disc, the extent of the disc, and the value of the
gravitational softening parameter (Plummer-softening). For the
purposes of this article, we have selected two models, hereafter
Model 1 and Model 2, which show the characteristics of the
straight segments most clearly, and discuss the other models only
briefly.

\begin{table}
\centering
 \caption{Essential  parameters in Models 1 and
   2. $M_\mathrm{disc}$ is the disc mass, $M_\mathrm{bulge}$ is the
   bulge mass, $M_\mathrm{halo}$ is the mass of the halo inside $R=15
   \ \mathrm{kpc}$, $b_{bulge}$  -- the bulge scale radius, and $R_C$ is
   the halo core radius. The initial value of the Toomre-parameter
   $Q_T$ and the gravitational softening $\epsilon$ are also
   indicated.}
  \begin{tabular}{lcc}
  \hline
   Model            &  1 & 2\\
\hline
$M_\mathrm{disc} [M_\odot]$ &  $2.9 \times 10^{10}$ & $3.0 \times 10^{10}$\\
$M_\mathrm{bulge} [M_\odot]$ & $9.2  \times 10^{9}$ & $1.5 \times 10^{10}$\\
$M_\mathrm{halo} [M_\odot], R < 15 \ \mathrm{kpc}$ & $9.9 \times 10^{10}$ & $1.1 \times 10^{11}$\\
$b_{bulge} [\mathrm{kpc}]$ & $0.6$ & $1.1$\\
$R_C$ & $7.5$ & $5.3$\\
$Q_T$ & $1.2$ & $1.1$\\
$\epsilon \mathrm[pc]$ & $75.0$ & $225.0$\\
\hline
\end{tabular}
\label{model_parameters}
\end{table}

\begin{figure}
\centering \resizebox{\hsize}{!}{\includegraphics{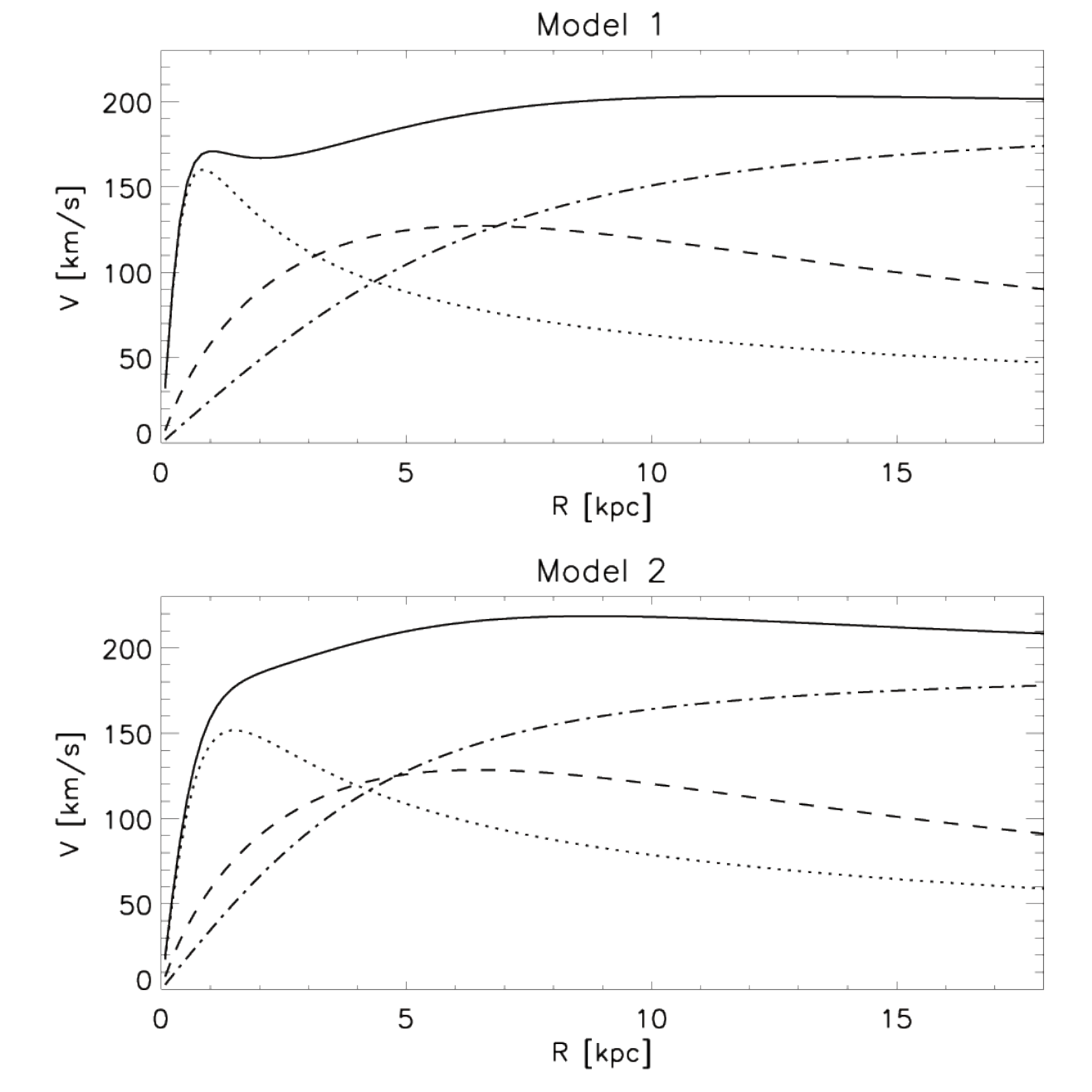}}
\caption{The rotation curves of models 1 and 2. The continuous
lines  show the total rotation curves, the bulge contribution is
drawn   with a dotted line, the disc contribution with a dashed
line,  and the  halo contribution with a dash-dotted line.}
\label{rotacurves}
\end{figure}

The rotation curves of Models 1 and 2 are shown in
Fig.~\ref{rotacurves}, with the adopted physical scaling of the
simulation units. In both models, the disc particles were
originally distributed as an exponential disc with scale length
$R_e= 3.0 \ \mathrm{kpc}$. The bulge component was modelled as an
analytical Plummer sphere, and the analytical halo was of the
same form as in \citet{rautiainen2010}. In both cases the
asymptotic rotation velocity of the halo is $189 \ \mathrm{km \
s}^{-1}$, but the core radius is different. As the rotation
curves show, both models are mostly dominated by the spherical
(analytical) component (bulge and halo); the reason for this
choice of parameters was to delay the bar formation, but the
initial value of the Toomre-parameter is low enough ($Q_T=1.2$ in
Model 1 and $1.1$ in Model 2)  that  still allows  the discs to
develop well-defined spiral arms. The essential model parameters
are given in Table~\ref{model_parameters}.

Model 1 first develops a  multi-armed structure. In the outer
parts of the disc there are $m=10\textrm{--}20$ short arms. The
structure becomes more regular in the inner parts. Even there the
number of arms  is varying ($m=2\textrm{--}5$). A large scale bar
forms at $T \approx 800 \ \mathrm{Myr}$. After its formation, the
inner spiral structure becomes effectively two-armed and the
number of spiral arms  diminishes also in the outer parts,
although  the outer spiral structure still remains multi-armed.

\begin{figure*}
\centering \resizebox{\hsize}{!}{\includegraphics{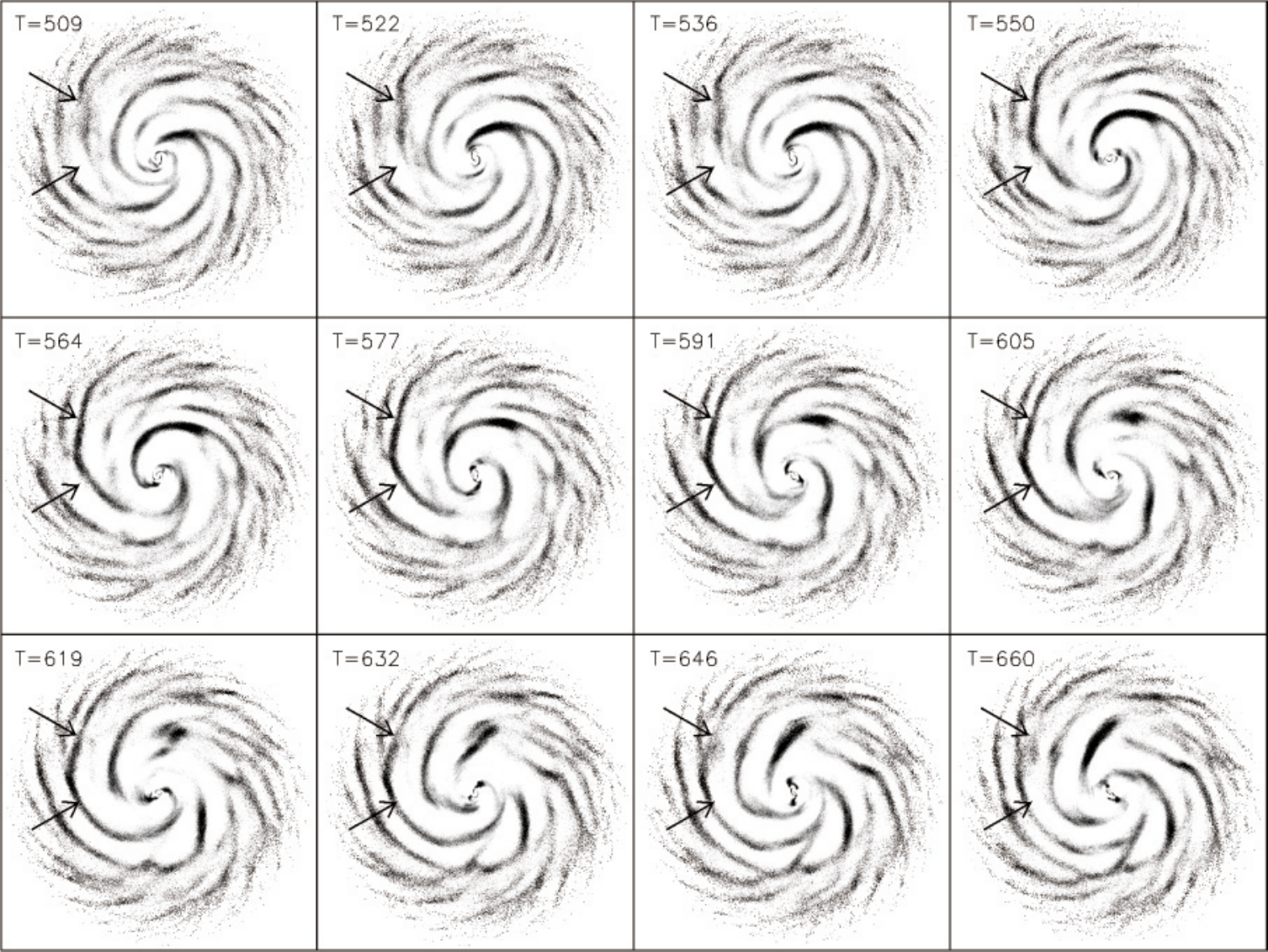}}
\caption{The formation and evolution of straight segments in one
  spiral arm. The frames have a width of 18 kpc and they show the
density enhancement above the average density at the same radius.
The time moment (in Myr) is exhibited at each frame. The two
arrows shown in the frames indicate the locations of two straight
segments. The densities are shown in a rotating coordinate system
(see details on text).} \label{polygon_evolution}
\end{figure*}

Fig.~\ref{polygon_evolution} demonstrates the evolution of two
straight segments (in one spiral arm), whose locations at $R
\approx 7\textrm{--}9.5 \ \mathrm{kpc}$ and $R \approx
9.5\textrm{--}13 \ \mathrm{kpc}$ are indicated with arrows. To
make following the evolution easier, we have used a rotating
coordinate system with angular velocity that keeps the two
straight segments in nearly the same place in our frames.  This
corresponds to pattern speed of about $18 \ \mathrm{km \ s}^{-1}
\mathrm{kpc}^{-1}$. In the beginning of the shown time sequence,
$T=509 \ \mathrm{Myr}$, the region $R=7\textrm{--}13 \
\mathrm{kpc}$ has $ m\approx10$ spiral arms.  However,
$10\textrm{--}30$ Myr later, the particles  form $m\approx 4$
longer spiral arms. One of them clearly has two straight
segments, forming  around $T=550 \ \mathrm{Myr}$  and  being
strongest at about $T=591 \ \mathrm{Myr}$. After  that, the
straight segments, and also the associated spiral arm itself,
become weaker. In the last couple of frames, there is no sign of
the two straight segments, but there is a new one in the opposite
side of the galaxy.

\begin{figure*}
\centering \resizebox{\hsize}{!}{\includegraphics{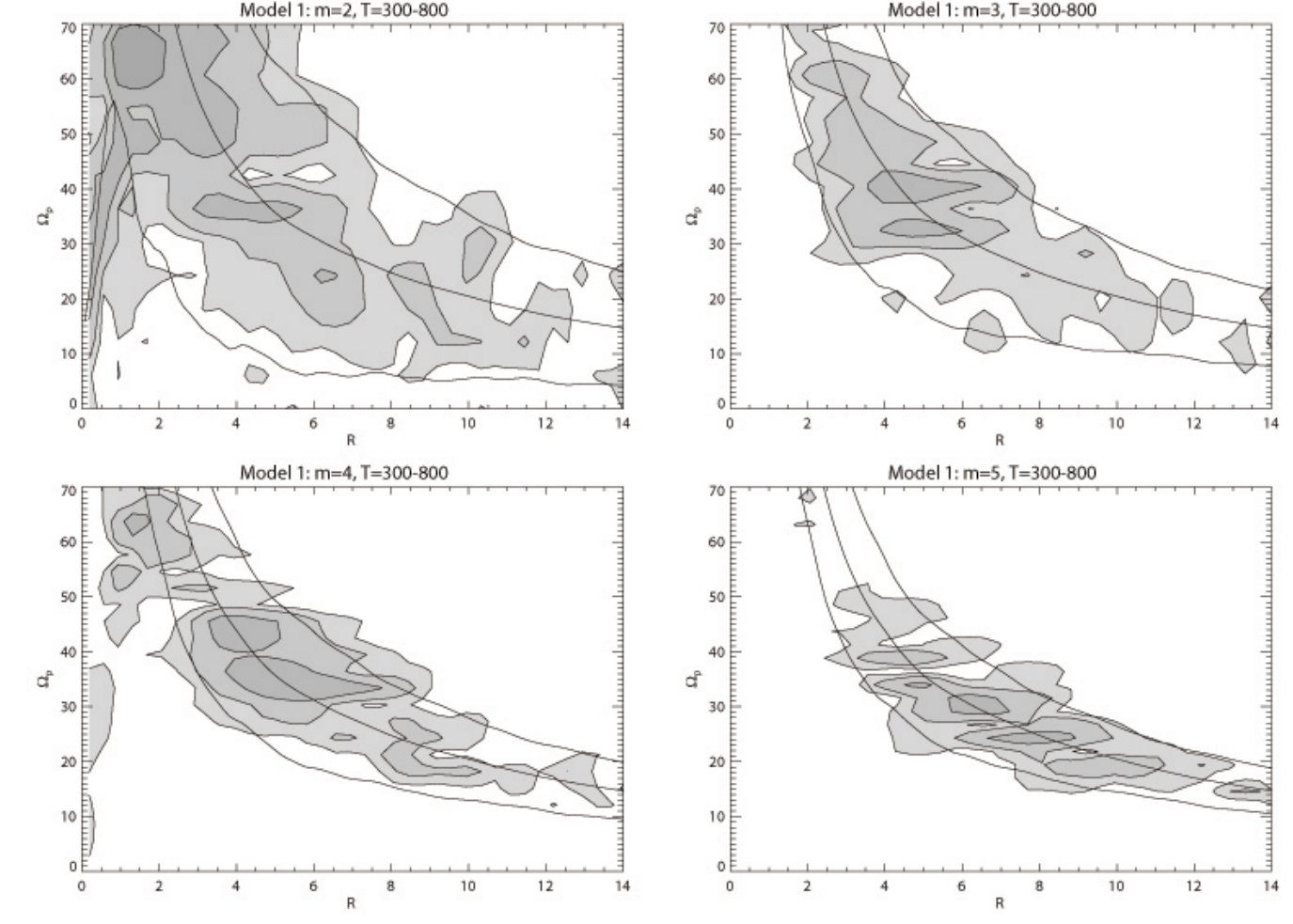}}
\caption{The amplitude spectra of  model 1 based on the Fourier
decomposition components $m=2\textrm{--}5$. The continuous lines
show $\Omega$ and $\Omega \pm \kappa/m$ in each frame.  The
contour levels are 0.025, 0.04, 0.1 and 0.2 above the azimuthal
average density at each radius.  The sampling period (in Myr) is
indicated in each frame.} \label{polygon_power}
\end{figure*}

These two segments in Model 1 were selected because they show
exceptionally well the formation and destruction of these
features. Also, they are exceptionally long-lived, lasting about
80 million years, which corresponds to about fourth of the
circular rotation period at the radial distance of the segments.
Most straight segments seen in our models have shorter lifespans,
corresponding to 10--30 millions of years.

More insight to the evolution of Model 1 can be obtained by
Fourier analysis of its surface density. Fig.~\ref{polygon_power}
shows the amplitude spectra \citep{masset1997,salo2000b} for the
$m=2\textrm{--}5$ components during the epochs when the two
straight segments appear. Also shown are the frequency curves
$\Omega$ and $\Omega \pm \kappa/m$. In the vicinity of the two
segments, i.e. $R=7\textrm{--}13 \ \mathrm{kpc}$, the strongest
feature can be seen in the $m=4$ and  in $m=5$ amplitude spectra.
This is not surprising, although the number of spiral arms in
this region is a bit varying, a four- or five-armed structure is
the most prevalent case.

The feature seen in the $R=7\textrm{--}13 \ \mathrm{kpc}$ region
both in $m=4$ and  in  $m=5$ spectra has a pattern speed
$\Omega_\mathrm{p} \approx 18 \ \mathrm{km \ s}^{-1}
\mathrm{kpc}^{-1}$, which corresponds to the corotation resonance
radius of $\sim 11 \ \mathrm{kpc}$, coinciding with outer of the
two segments.  In the $m=5$ spectrum, there is also a clear
feature with $\Omega_\mathrm{p} \approx 24\ \mathrm{km \ s}^{-1}
\mathrm{kpc}^{-1}$, which probably has an effect on the inner
segment. There are also features in the spectra of higher values
of $m$ (6--12), but these are limited to immediate vicinity of
the $\Omega$-curve.

In Model 2 the disc does not form a large scale bar during the
simulation time, which corresponds to about 5 Gyr. The disc shows
mostly multi-armed ($m=5\textrm{--}10$) morphology, which
occasionally develops straight segments. In many time steps these
arms appear to be long, extending throughout most of the disc,
but a closer look at their evolution and the amplitude spectra
demonstrates that they actually consist of  a large number of
short features, whose pattern speeds are close to  the local
circular velocity. This kind of behavior resembles the recent
models by \citet{grand2012} and \citet{d_onghia2013}. In the
later phase of the simulation, the number of arms in the inner
parts of the disc diminishes to $m=2\textrm{--}4$, and the
innermost part resembles a small bar or oval.

In other models, which are not shown or analysed in this article,
we made further experiments with model parameters, such as the
value of the gravitational softening and the initial extent of
the stellar disc. There is already a quite large difference in
softening parameters between models 1 and 2, and tests with other
values show that it is not critical for the formation of straight
segments, as long as its value is not so high  to suppress the
formation of all the sharp features on the disc. The situation is
quite similar with the disc extent; there are more straight
segments in larger discs, but even models, where the initial
stellar particle distribution reaches only 6 kpc or two disc
scale lengths, can have them.

\section{General characteristics of straight segments in model discs}

We  identified straight segments in stellar discs of models 1 and
2. For their identification we have used the images of model
discs processed by the masking program, which leaves only regions
of  enhanced density, i. e. regions where the density exceeds its
average level at the same radius. This procedure increases the
contrast between the arms and the inter-arm space and facilitates
the study of the galactic morphology. In identification of
straight segments we adhered to the following principles.

\begin{enumerate}

\item The line, connecting the ends of a straight segment, must
lie entirely  in the region of the enhanced density.

\item The ends of straight segments must have some specific
features: either the density dropping  below the average level or
the significant increase in the pitch angle of a spiral arm.

\item In all cases we try to identify straight segments so that
their length $L$ would be maximal.

\item The straight segment must be quite elongated: the ratio of
its length to the width must exceed 4.

\end{enumerate}

Table 1 exhibits the average characteristics of the straight
segments identified in models 1 and 2. It shows the total number
$N$ of moments considered, the number of the selected straight
segments $n_s$, the coefficient $k$ in the dependence $L=kR$, and
its error. It also includes the standard deviations $\sigma_0$
and $\sigma_1$ calculated for linear  relation $L=kR$ and for
non-linear law $L=2.4f_d(R)R$, respectively. We also present the
average value of the angle $\overline{\beta}$ between the
straight segment and the azimuthal direction, its standard
deviation (in brackets), the average value of the angle
$\overline{\alpha}$ between two neighboring straight segments,
its standard deviation (in brackets), and the number $n_\alpha$
of measurements of $\alpha$.

\begin{table}
\centering
 \caption{Characteristics of the model straight segments}
  \begin{tabular}{lcc}
  \hline
   Model            &  1 & 2\\
 \hline
$N$ moments         &  36& 53\\
$n_s$                 &  238& 273\\
$k$ in $L=kR$       &  $0.86\pm0.02$& $0.88\pm0.01$\\
$\sigma_0$ in $L=kR$ &  2.10 kpc & 1.80 kpc\\
$\sigma_1$ in $L=2.4f_dR$ &  1.54 kpc &  1.51 kpc \\
$\overline{\beta}$  &  $28^\circ$ ($9^\circ$)& $25^\circ$ ($9^\circ$)\\
$\overline{\alpha}$ &  $127^\circ$ ($13^\circ$)& $125^\circ$ ($11^\circ$)\\
$n_\alpha$          &  101& 129\\
\hline
\end{tabular}
\end{table}


\subsection{$L$--$R$ dependence}

\begin{figure*}
\centering \resizebox{\hsize}{!}{\includegraphics{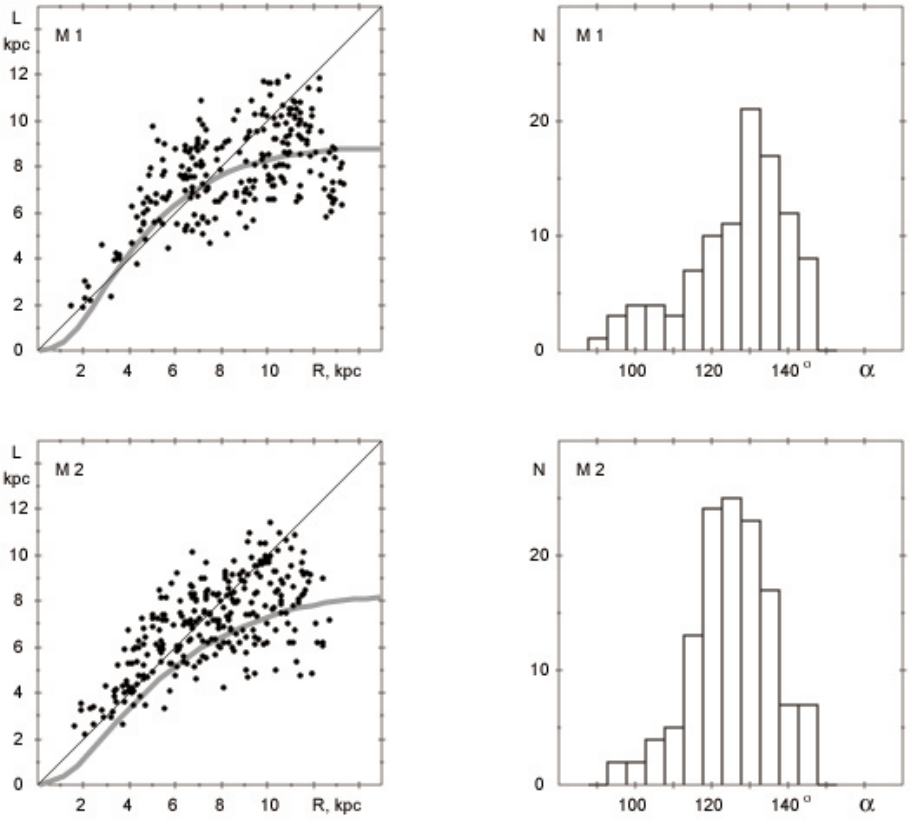}}
\caption{Left panel: the dependence between the length $L$ of a
straight segment and its Galactocentric distance $R$ in models 1
and 2. The thick gray curve shows the value of $L$ calculated
from the formula $L=2.4 f_d(R) R$. The bisectrix is also drawn.
Right panel: the histograms of  the distribution of the angle
$\alpha$ between two neighboring straight segments.}
\label{L-R+his}
\end{figure*}

\begin{figure}
\centering \resizebox{\hsize}{!}{\includegraphics{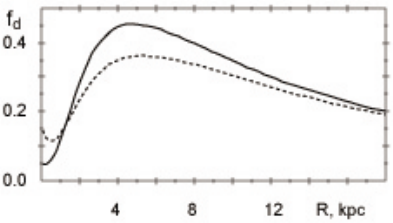}}
\caption{Variations in  $f_d$, the relative contribution of the
disc to the total rotation curve ($v_c^2$),  in models 1 (solid
line) and 2 (dashed line).} \label{fd}
\end{figure}

The variations in the length of the straight segments $L$ along
radius $R$ in models 1 and 2 are shown in Fig.~\ref{L-R+his}. The
Galactocentic distance $R$ for the straight segment is determined
as the distance to its  median point. The thick gray curve shows
the value of $L$ calculated from the formula $L=2.4 f_d R$, where
$f_d(R)$ is the relative contribution of the disc to the total
rotation curve at each radius. The value of $f_d(R)$ achieves the
maximum  $R\approx 5$ kpc and then gradually decreases with
increasing $R$ (Fig.~\ref{fd}).

We can see that both dependencies $L=(0.86\pm0.02)R$ (model 1)
and $L=(0.88\pm0.01)R$ (model 2) derived  for the model straight
segments are consistent with observations, $L=(1.0\pm 0.13) R$.
However,  the connection between $L$ and $R$ is conspicuously
non-linear in both models: there are  a lot of relatively short
straight segments at  large radii. So  $L$--$R$ relation  is
better described by formula $L=2.4 f_d R$. The standard deviation
$\sigma_1$ is less than $\sigma_0$ derived for the linear law by
27 per cent in model 1 and by 16 per cent in model 2. This
difference can be related to the fact that the amplitude of
variations of $f_d$ is larger in model 1 than in 2
(Fig.~\ref{fd}).


\subsection{The angle between the neighboring segments}

Since the straight segments rotate  in the disc  with  the
angular velocity of their parent overdensities, they can never
form stationary polygonal structure. Moreover, straight segments
must destroy each other during their merging. The only possible
way for their contact without destruction is a touch   with their
edges. In this case they can even increase each other, because
the appearance of an extra density at their endpoints gives both
of them an extra ability to hold stars inside them.

Table 1 indicates that the average value of the angle $\alpha$
between two neighboring straight segments is
$\overline{\alpha}=127^\circ$  and $\overline{\alpha}=125^\circ$
in models 1 and 2, respectively.  The maximum in distribution of
$\alpha$ lies near $\alpha=130^\circ$ in both cases
(Fig.~\ref{L-R+his}). All these values are consistent with
observations.

\citet{chernin1999} gives an explanation of the  value of
$\overline{\alpha}=120^\circ$, which is  based on the relation
$L=R$. We present here a bit different explanation, which also
invokes the correlation between $L$ and $R$.

\begin{figure*}
\centering \resizebox{15 cm}{!}{\includegraphics{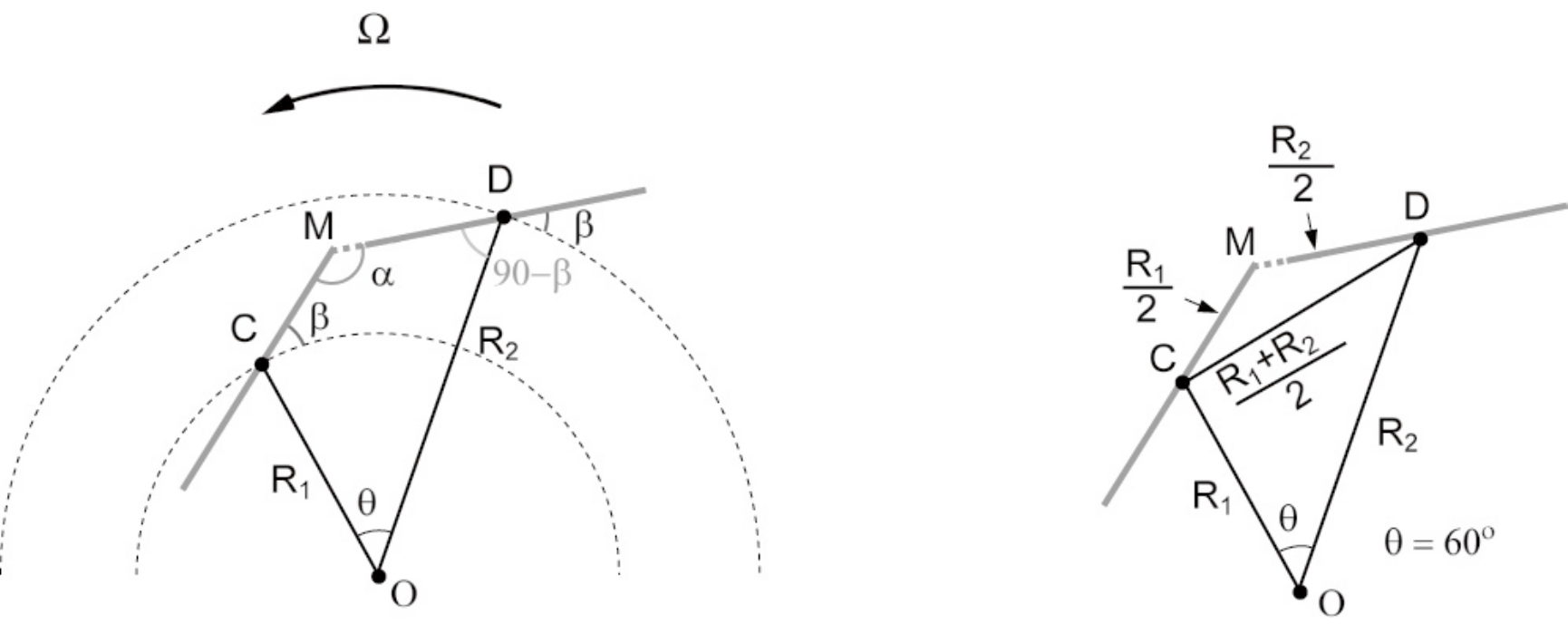}}
\caption{Two straight segments (gray lines) at the moment of
contact. The angle $\alpha$ between  them can be found from the
quadrangle OCMD:
$\alpha=360^\circ-\theta-(90^\circ+\beta)-(90^\circ-\beta)=180^\circ-\theta$.
As the triangle COD has sides $R_1$, $R_2$, and $\sim
(R_1+R_2)/2$,  the angle $\theta$  equals $\theta\approx
60^\circ$. Consequently, angle  $\alpha$ is practically
independent of $\beta$ and equals $\alpha \approx 120^\circ$.}
\label{triangles}
\end{figure*}

Let us consider two straight segments at the moment, when they
touch each other with their edges (Fig.~\ref{triangles}). They,
together with the radius-vectors, form the quadrangle OCMD. We
are looking for the angle $\alpha$ between two straight segments.
As the sum of the angles in a quadrangle  is 360$^\circ$, we can
find the angle $\alpha$ by subtracting the other angles from this
value. The galactocentric angle is denoted by $\theta$, two other
angles have values $90^\circ+\beta$ and $90^\circ-\beta$. So it
is the angle $\theta$ that determines the value of $\alpha$:
$\alpha=180^\circ-\theta$. We can find $\theta$ from the triangle
COD.  Due to the relation $L=R$ the side CD in the triangle COD
has the value $\sim (R_1+R_2)/2$, what correspond to
$\theta\approx60^\circ$ (use the law of cosines and neglect terms
of $(R_1-R_2)^2/R_1R_2$). Thus, the angle $\alpha$  has the value
of $\alpha=120^\circ$ and is  practically independent of $\beta$.

We have found that  in  models 1 and 2 the  coefficient $k$  in
the relation $L=kR$ is less than unity, $k=0.86\textrm{--}0.88$,
so the angle $\theta$ must be less than $\theta<60^\circ$ here.
Under $k=0.87$ it must have the value of $\theta=52^\circ$ and,
consequently, $\alpha$ must be
$\alpha=180^\circ-\theta=128^\circ$. The last value is in  good
agreement with corresponding estimates in our models.

\subsection{The angle $\beta$ between  the straight segments and the azimuthal direction}

We measured the angle $\varphi$ between the straight segment and
the radius-vector drawn from the galactic centre to the median
point of the straight segment and calculated the angle $\beta$,
supplementing it up to 90$^\circ$, $\beta=90-\varphi$. Generally,
angle $\beta$ is an analog of the pitch angle for the spiral
arms. Its average value equals $28^\circ$ and $25^\circ$ in
models 1 and 2, respectively (Table 1).

Fig.~\ref{beta} shows the variations of $\beta$ along radius and
the histograms of the distribution of $\beta$. We can see that
the angle $\beta$, on average, decreases with radius. The
approximation of these variations by the linear law gives the
following parameters: $\beta=(-1.63\pm0.17)R+41.7\pm1.5$ for
model 1 and $\beta=(-1.68\pm0.17)R+37.6\pm1.3$ for model 2, where
$\beta$ is in degrees and $R$ in kpc.

The variations in $\beta$ along $R$ can be partly (within
10$^\circ$) explained by the deviations of the model rotation
curves from  flat one. For  non-flat rotation curve  the angle
$\beta$ can be estimated from the relation:

\begin{equation}
  \beta\approx\arctan\frac{\kappa}{\pi R|\Omega'(R)|},
  \label{beta}
\end{equation}

\noindent where $\Omega'(R)$ is the first derivative of
$\Omega(R)$ with respect to $R$. This expression is combination
of Eqs.~2 and 3, but obtained for the case "with self-gravity",
in which the maximal density corresponds to the point "3"
situated one-half of the epicyclic period ($\pi/\kappa$)
downstream the initial disturber (Fig.~\ref{trajectory-1}). In
the case of flat rotation curve, Eq. 12 transforms to Eq.~6.
Generally, the rising rotation curve  increases $\beta$, while
the descending one decreases it.

\begin{figure*}
\centering \resizebox{\hsize}{!}{\includegraphics{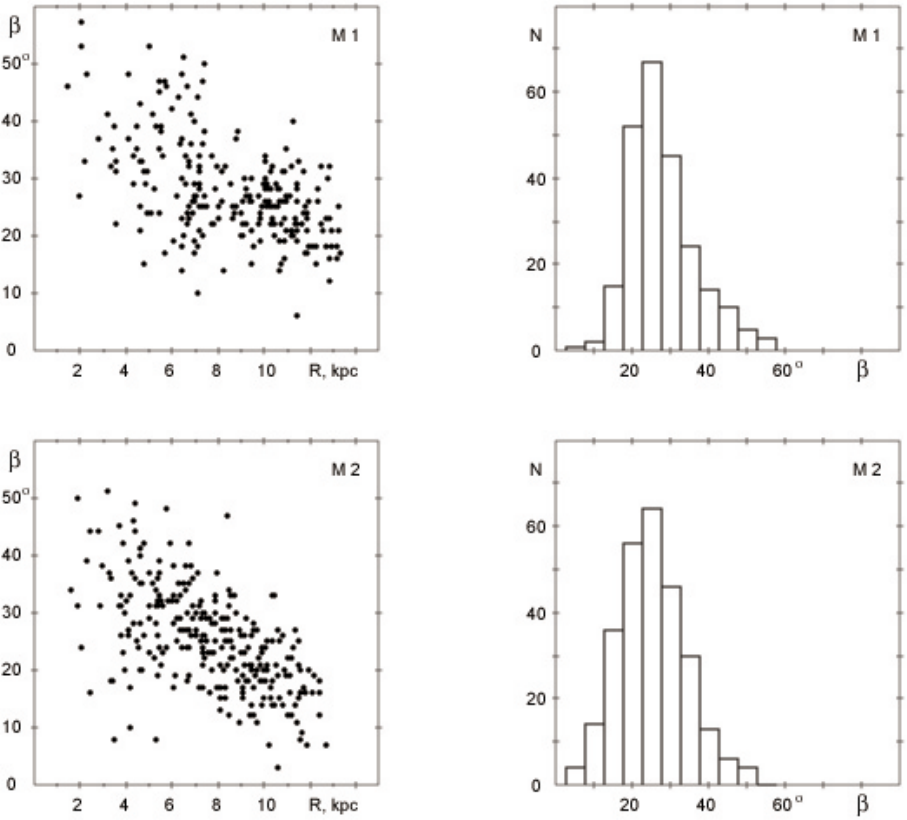}}
\caption{Left panel: variations in the   angle $\beta$ (one
between the straight segment and the azimuthal direction) along
the radius $R$ in models 1 and 2.  Right panel: the histograms of
distribution of $\beta$.} \label{beta}
\end{figure*}

\section{Kinematical features of the straight segments}

The role of the initial overdensities  in production of the
straight segments is to adjust the epicyclic motions of stars
passing by. So the overdensities must create the specific
velocity field in their neighborhood. To study the kinematics of
stars in our models we calculated the residual velocities of
stellar particles in the radial and azimuthal directions, $V_R$
and $V_T$.

In our previous paper \citep{rautiainen2010} we determined $V_R$
and $V_T$ as differences between the model velocities and the
velocity due to  rotation curve, but there we considered the gas
subsystem, which rotated practically with the velocity of
rotation curve. However, it is not true for the stellar discs.
Due to the conspicuous velocity dispersion, the stellar discs
rotate, on average, with the smaller velocity than that of
rotation curve. It is so-called asymmetric drift
\citep{binney2008}. In the present paper we calculate the
azimuthal residual velocity $V_T$ with respect to the average
azimuthal velocity of stellar particles at the same radius, but
not with respect to the rotation curve. Nothing have changed for
the radial residual velocity $V_R$, which  coincides with the
radial velocity  with respect to the origin.

\begin{figure*}
\centering \resizebox{\hsize}{!}{\includegraphics{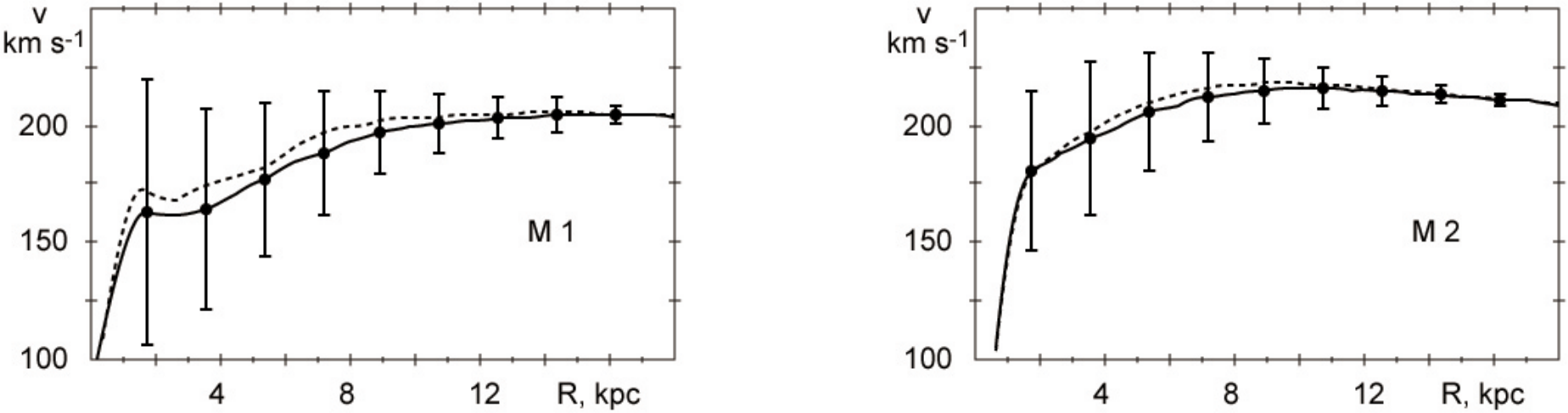}}
\caption{The average azimuthal velocity of stars
$\overline{v}_\theta$ (solid line) and  the velocity of rotation
curve $v_c$ (dashed line) in models 1 and 2. The bars represent
the velocity dispersion $\sigma_R$. Calculations are made for
moments $T=632.5$ and $T=1402.5$ Myr in models 1 and 2,
respectively.} \label{dispersion}
\end{figure*}

Fig.~\ref{dispersion} demonstrates the distribution of the
average azimuthal velocity of stars $\overline{v}_\theta$ and
that of the rotation curve $v_c$ along radius in  models 1 and 2.
It also shows the velocity dispersion in radial direction
$\sigma_R$ at different radii.  For example, at $R=7$ kpc, the
asymmetric drift amounts $v_c-\overline{v}_\theta=9$ and 4 km
s$^{-1}$  in models 1 and 2, respectively. And the velocity
dispersion $\sigma_R$ at the same distance  has the value of 26
and 20 km s$^{-1}$ in models 1 and 2, respectively. Generally,
the asymmetric drift and the velocity dispersion are larger in
model 1.

Let us consider the velocity field created  by the overdensities,
corotating with the disc, in two cases: without self-gravity  and
with it.

Without self-gravity, the maximal density of the straight segment
must correspond to the maximal absolute value of the azimuthal
residual velocity $V_T$. At the radii larger than that of the
initial disturber, $V_T$ must be directed in the sense of the
galactic rotation (Fig.~\ref{trajectory-1}a), so it must have
positive value ($V_T>0$ under $R>R_0$), while at the smaller
radii it must be directed in the opposite sense ($V_T<0$ under
$R<R_0$). As for the radial velocity, stars achieve their maximal
absolute value of  $V_R$, when they are leaving  the straight
segment. Thus, without self-gravity, stars in regions of enhanced
density must have conspicuous velocity $V_T$ and nearly zero
$V_R$.

When  self-gravity is important,  the maximal density of the
straight segment must correspond to the maximal radial velocity
$V_R$: at radii larger than that of the initial disturber, $V_R$
must be directed away from the galactic centre and be positive
($V_R>0$ under $R>R_0$) (Fig.~\ref{trajectory-1}b), while at the
smaller radii it must be directed toward the galactic centre
($V_R<0$ under $R<R_0$). The azimuthal velocities, on the
contrary, achieve their extremal values, when stars  leave the
straight segment. So with self-gravity, regions of enhanced
density must exhibit conspicuous velocity $V_R$ and nearly zero
$V_T$.

It is possible a mixed case, when both the gravity of an initial
disturber and self-gravity of a straight segment are important.
In this case, we can observe the conspicuous gradient of the
radial and azimuthal velocities in the straight segments near
overdensities. But both gradients must have definite direction:
the larger (smaller) $R$ the more positive (negative) values of
$V_R$ or $V_T$.

To study the kinematics in the model discs, we divided them into
small squares with the size of $150\times 150$ pc and  calculated
the average  radial and azimuthal residual velocities, $V_R$ and
$V_T$,  for stars located inside them at different moments. We
divided the  velocities into tree groups: negative, positive, and
close to zero, the latter were those, which didn't exceed 3 km
s$^{-1}$ in  absolute value, $|V_R|<3$ or $|V_T|<3$ km s$^{-1}$.

\begin{figure*}
\centering \resizebox{\hsize}{!}{\includegraphics{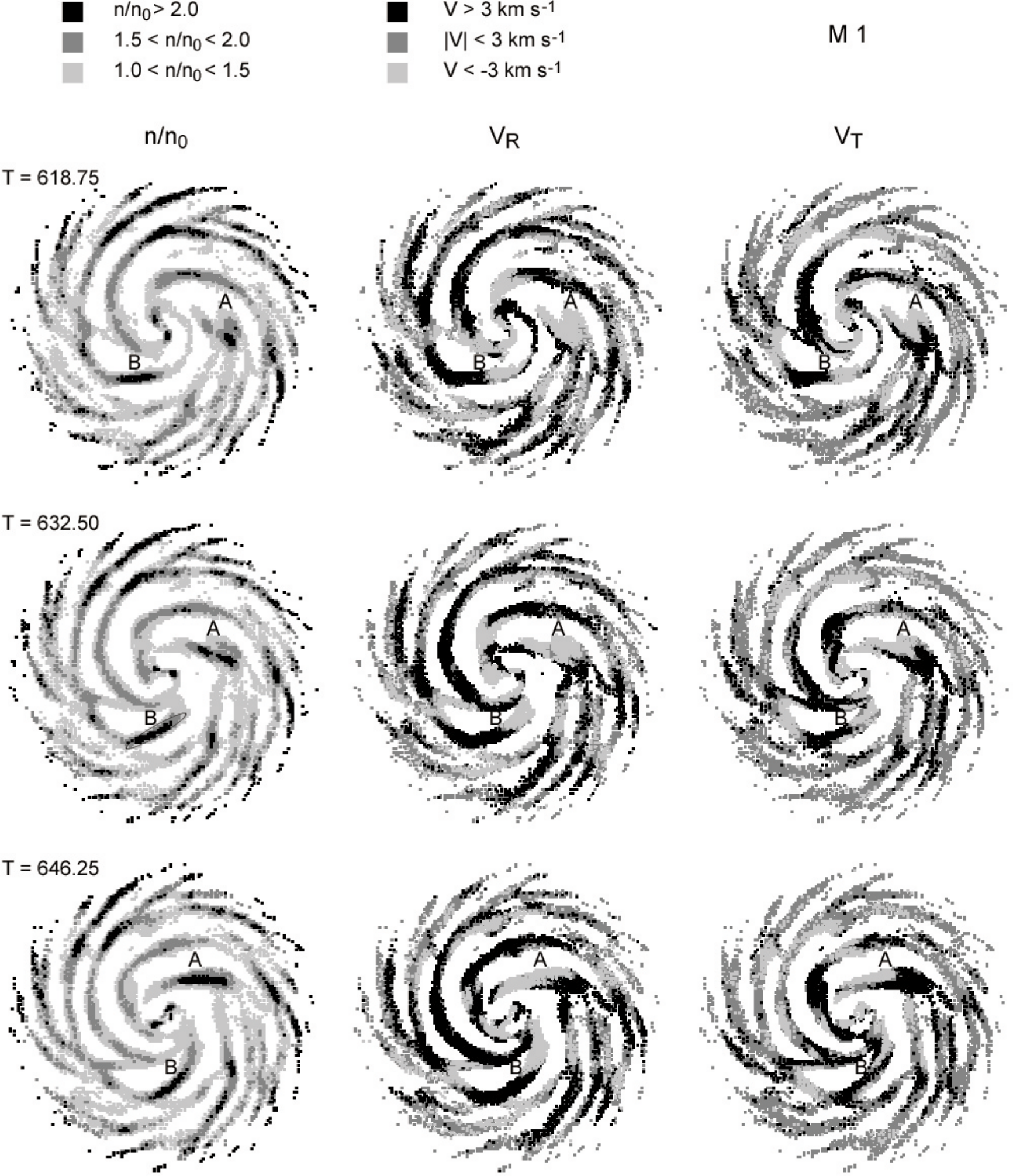}}
\caption{Distribution of the radial $V_R$ and azimuthal $V_T$
residual velocities averaged in squares $150\times 150$ pc
throughout the galactic disc in model 1 at three moments
$T=618.75$, 632.50, and 646.25 Myr (tree rows). The average
velocities are divided into tree groups: negative (light gray
squares), close to zero (dark gray squares), and positive ones
(black squares). The first column shows the distribution of the
relative density $n/n_0$ in the galactic disc, where $n$ -- the
number of particles in a square and $n_0$ -- the average number
of particles in squares at the same radius. The greater the
density the darker the color of the square. Two overdensities are
designated by letters "A" and "B". Near them the velocities $V_R$
and $V_T$ demonstrate the following gradients: the larger
(smaller)  $R$ the more positive (negative) velocity.}
\label{velocities_1261}
\end{figure*}

Fig.~\ref{velocities_1261} exhibits the distribution of the
radial $V_R$ and azimuthal $V_T$ velocities averaged in squares
in model 1 at three moments $T=618.75$, 632.50, and 646.25 Myr
(tree rows). Positive velocities ($V_R$ or $V_T$) are shown in
black and the negative ones -- in light gray, the velocities
close to zero are denoted in dark gray. The first column shows
the distribution of the relative density $n/n_0$ in the galactic
disc, where $n$ is the number of particles in a square and $n_0$
-- the average number of particles in squares at the same radius.
The greater the density the darker the color of the square.

We can follow the formation of the straight segments near two
overdensities designated by letters "A" and "B". There are
conspicuous gradients of velocities $V_R$ and $V_T$ near them at
all three moments. And the  directions of these gradients
coincide with the expected ones.

\begin{figure*}
\centering \resizebox{\hsize}{!}{\includegraphics{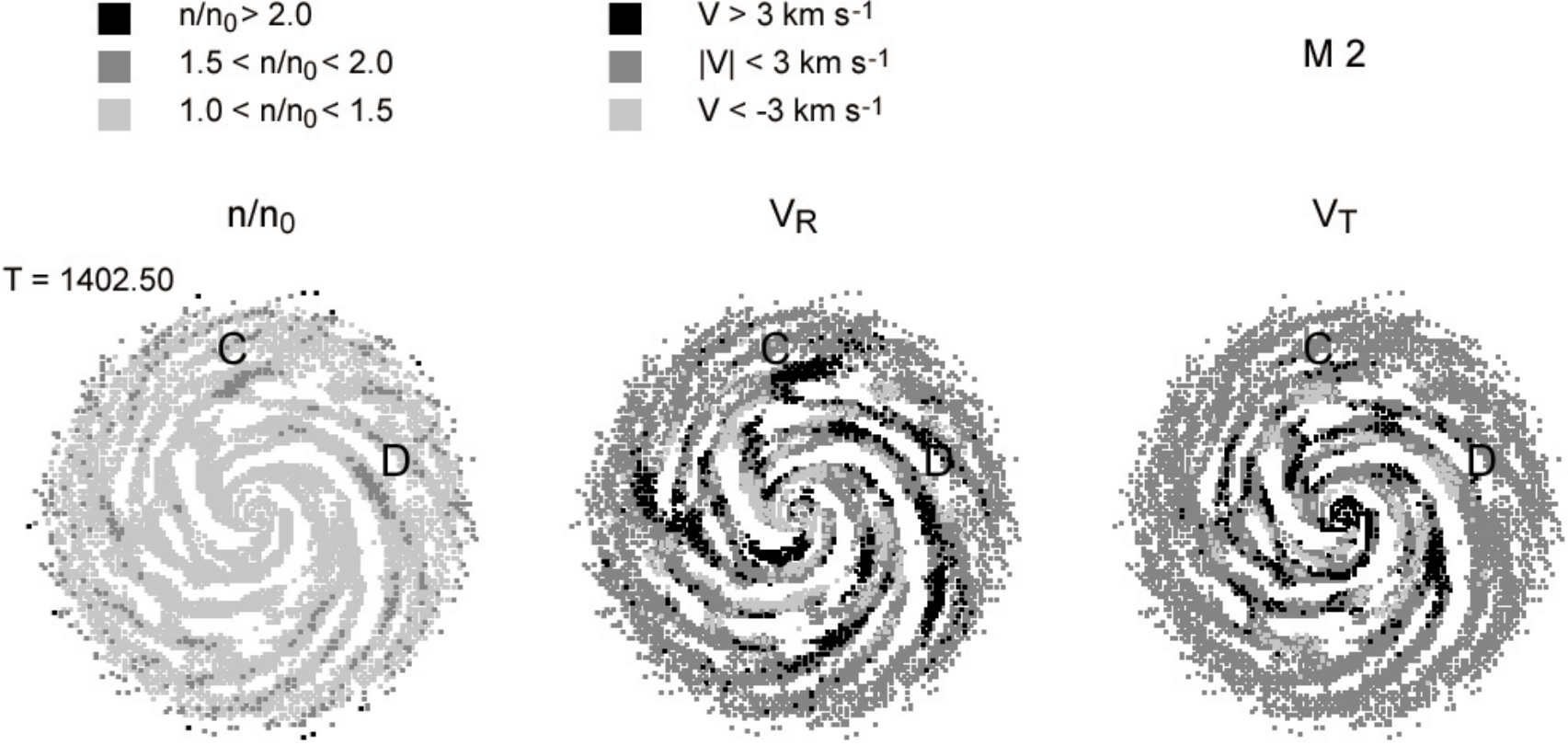}}
\caption{Distribution of the residual velocities $V_R$ and $V_T$
averaged in squares throughout the galactic disc in model 2 at
$T=1402.50$ Myr. The left image shows the distribution of the
relative density. For more details see caption to
Fig.~\ref{velocities_1261}. Two over-densities are designated by
letters "C" and "D". Near them the velocities increases
(decreases) with increasing (decreasing) $R$.  Model 2 exhibits
density and velocity perturbations of less amplitude in
comparison with model 1.} \label{velocities_205}
\end{figure*}

We also study the velocity field in model 2.
Fig.~\ref{velocities_205} shows the distribution of the relative
density and residual velocities, $V_R$ and  $V_T$,  averaged in
squares  throughout the galactic disc in model 2 at $T=1402.50$
Myr. Two overdensities are designated by letters "C" and "D". We
can see the expected velocity gradients near them as well.

\begin{figure}
\centering \resizebox{\hsize}{!}{\includegraphics{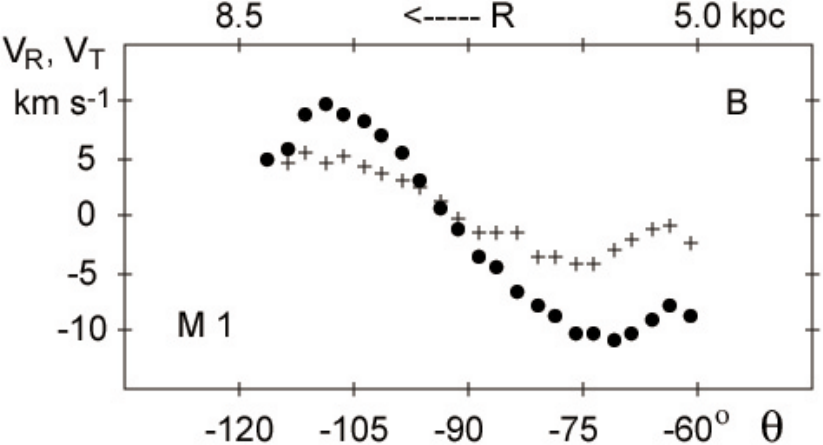}}
\caption{The radial $V_R$ (black circles) and azimuthal $V_T$
(crosses) residual velocities of stars located inside detail "B".
The velocities were calculated in sectors of width $\Delta
\theta=2.5^\circ$ along galactocentric angle $\theta$. The range
of changes in $R$ is shown at the upper boundary. }
\label{gradients}
\end{figure}

However, Figs.~\ref{velocities_1261} and ~\ref{velocities_205}
demonstrate the direction of the velocity gradients but not the
amount of velocity changes. To illustrate them we selected stars
inside  detail "B". Specifically, we  took out 101084 stars
located inside ellipse shown in  Fig.~\ref{velocities_1261} at
$T=632.50$ Myr (1-st column). The stars were divided into sectors
of width $\Delta \theta=2.5^\circ$ along the galactocentric angle
$\theta$. In each sector we calculated the average radial $V_R$
and azimuthal $V_T$ residual velocities, which are shown in
Fig.~\ref{gradients}. We can see that the range of changes of the
velocity $V_R$ is $\pm 10$ km s$^{-1}$ and that of $V_T$ is $\pm
5$ km s$^{-1}$. Note that the geometry of pieces of trailing
spiral arms is such that the increase in $\theta$ corresponds to
the decrease in $R$. The range of changes in $R$ is shown at the
upper boundary of Fig.~\ref{gradients}. For  comparison, the
range of changes the velocities $V_R$ in detail A ($T=646.25$
Myr) is $\pm 7$ km s$^{-1}$, but that in details C and D is from
-1 to +5  km s$^{-1}$.

On the whole, the distribution of the negative and positive
residual velocities  agrees with hypothesis that the straight
segments are forming as the response of the disc to the
overdensity corotating with it. The amplitude  of velocity
changes varies from a few  to 10 km s$^{-1}$. Generally, model 2
exhibits density and velocity perturbations of less amplitude in
comparison with model 1.


\section{Discussion and conclusions}

We consider the formation of the straight segments in the stellar
galactic discs. For this purpose we constructed two N-body
simulations, which differs in concentration of mass to the
galactic centre. In  model 1 the stellar disc forms the bar in
the central region, while in model 2 the central part of the disc
is occupied by the multi-armed spiral pattern.

We  identified more than 500 straight segments in the two models.
The straight segments  are temporal features, which rotate with
the average velocity of the disc. The relation between the length
$L$ of the model straight segment and its Galactocentric distance
$R$ can be approximated by the linear law $L=kR$ with the
coefficients lying in the range $k=0.86\textrm{--}0.88$. The
average angle between two neighboring straight segments in our
models appears to be $\overline{\alpha}=125
\textrm{--}127^\circ$. All these values are consistent with the
observational estimates, $L=(1.0\pm0.13)R$ and
$\alpha=120^\circ$, derived  by \citet{chernin2001}.

We suggest that the formation of the straight segments in stellar
discs is connected with the appearance  of   overdensities
corotating with the disc. The great role of such overdensities is
revealed in the numerical experiments by \citet{d_onghia2013}. In
the first approximation, the response of the stellar disc to such
overdensity  must have the shape of a straight segment with the
length determined by the formula $L=2.4 f_d R$.

Comparison of the average characteristics of the model straight
segments with  the parameters of the respondent perturbations
shows that the non-linear law $L=2.4 f_d R$ describes better the
connection between $L$ and $R$ than the linear one $L=kR$
(Fig.~\ref{L-R+his}, Table 1).

We suppose that the straight segments can form the polygonal
structures  only when they touch each other  by their edges. In
other cases they must destroy each other. Using this hypothesis,
we can explain, why  the average value of the angle $\alpha$
between two neighboring segments  appears to be
$\overline{\alpha}=125\textrm{--}127^\circ$ in our models.

The angle $\beta$ between the straight segment and the azimuthal
direction  has the average value of
$\overline{\beta}=25\textrm{--}28^\circ$ in our models.  We found
the conspicuous decrease in $\beta$ with radius, that can be only
partly (within 10$^\circ$) related to the deviations of the model
rotation curves from  flat one. Fig.~\ref{beta} exhibits
relatively large values of $\beta$ in the central and
intermediate regions ($R<6$ kpc) in both models. One possible
explanation of these  departures is that the bar or oval modes
can interfere directly in the formation of the straight segments
here.

We study the kinematics of stars near the overdensities forming
in the stellar discs. For this aim we divided model discs into
small squares, $150\times150$ pc, and calculated average residual
velocities  in the radial and azimuthal directions, $V_R$ and
$V_T$. We found specific velocity gradients in the straight
segments near the overdensities: at the radii larger than that of
the overdensity, the velocities $V_R$ and $V_T$ are positive,
while at the smaller radii they are negative. Such velocity field
agrees with the hypothesis that the straight segments are forming
due to the tuning of the epicyclic motions near the initial
disturbers. The amplitude  of velocity changes inside straight
segments can achieve  10 km s$^{-1}$.

The most interesting question is the nature of the overdensities
bringing the formation of the straight segments. We suppose that
the appearance of such overdensities in our models is connected
with the interaction of different modes or waves, forming on the
galactic periphery and in more central region of the disc. This
suggestion has some kinematical foundation. The stars located in
the spiral arms inside and outside the CR have opposite phase of
the epicyclic motions, and consequently, the opposite residual
velocities. Probably, the superposition of such waves destroy the
adjusted epicyclic motions of both waves and create
overdensities, which have no systematic residual velocities and
nearly corotate with the disc.

\section*{\rm \large ACKNOWLEDGMENTS}

We thank H. Salo for using his N-body code. This work made use of
data from the Ohio State University Bright Spiral Galaxy Survey,
which was funded by grants AST-9217716 and AST-9617006 from the
United States National Science Foundation, with additional
support from the Ohio State University. The present work was
partly supported  by the Russian Foundation for Basic Research
(project nos.~12\mbox{-}02\mbox{-}00827,
13\mbox{-}02\mbox{-}00203).

\end{document}